\documentclass{cs21proc}

\usepackage{kantlipsum}

\editors{A. S. Brun, J. Bouvier, P. Petit}
\publisher{Zenodo}
\conference{The 21th Cambridge Workshop on Cool Stars, Stellar Systems, and the Sun}
\conferencedate{2022}

\title{Numerical simulation of convection in K dwarfs}
\author{Manfred K\"uker}

\affiliation{Leibniz Institute of Astrophysics Potsdam, An der Sternwarte 16, 14482 Potsdam, Germany}

\shorttitle{Convection in K dwarfs}
\shortauthors{M. K\"uker}

\abs{We have carried out numerical simulations of the convection zone in a K dwarf of 0.7 solar masses rotating at the solar rotation period. We study the convection pattern, the differential rotation and meridional flows, and the dynamo-generated magnetic field. We find that for a star of this type, the solar rotation period represents a case of fairly rapid rotation and the differential is solar-type.  A dynamo-generated large scale field appears but it is neither dipolar nor does it show a simple activity cycle.}

\begin{document}

\maketitle

\section{Introduction}
Magnetic activity is a common phenomenon for cool stars. A common property of these stars is an outer convection zone that covers a significant fraction of the stellar radius. While these convection zones contribute little to the total stellar mass, they comprise a large fraction of the stellar volume. To this day, there is no generally accepted theory for the generation of stellar magnetic fields. Convection, however, is usually a vital ingredient in current models. As the star rotates, the convective gas motions are affected by the Coriolis force, which causes them to be helical. It also causes a transport of angular momentum through Reynolds stress, which in turn causes differential rotation. The latter is a powerful generator of toroidal (azimuthal) magnetic fields, provided a poloidal field is already present. Helicity, on the other hand, can produce poloidal as well as toroidal fields. The best known dynamo model is the $\alpha \Omega$ dynamo, a combination of differential rotation and helicity.

While mean field models have historically had some success explaining the existence of stellar magnetic fields, they can not reproduce all the properties of stellar and particularly solar activity. Progress in computer technology over the last decades has made direct numerical simulations feasible. Starting from early work in two dimensions \citep{Gilman1981, Glatzmaier1981a}, a large variety of models have been developed in three-dimensional spherical geometry, covering either full spherical shells of wedge geometry. We here use the anelastic approach in full spherical shell geometry, which was first used by \cite{Glatzmaier1984}. Extensive use has been made of it since in the context of rotating stellar convection and magnetoconvection, particularly with the {\sc ASH} and {\sc Magic} codes \citep{Miesch2000, Brun2004, Christensen2001}. 
%
%
%
%
\section{Numerical simulations}

We use the Rayleigh MHD code \citep{Featherstone2016} in anelastic mode to simulate the gas flow, heat flux, and magnetic field in a spherical shell. The anelastic approximation allows larger time steps than a fully compressible code and is valid for small Mach numbers. 

To run the code for a particular stellar model, a background model is required to implement the stratification in the stellar convection zone. 
We use the {\sc Mesa} stellar evolution code \citep{Paxton2011, Paxton2013, Paxton2015, Paxton2018, Paxton2019} to produce a model for a main sequence star of 0.7 solar masses and solar metallicity and an age of 5 Gyr. With an effective temperature of 4368 K and a radius of $0.67 R_\odot$, the star has a luminosity of $0.15 L_\odot$. The bottom of the convection zone is at a radius of $0.45 R_\odot$.
 
To incorporate the stellar model into the Rayleigh MHD code, we approximate it with a simple polytropic model by solving a Lane-Emden-type system of ordinary differential equations for mass density, temperature, and gravity with the values from the {\sc Mesa} code at the inner boundary as initial conditions. Our model differs from that built into the Rayleigh code as it does not assume a $1/r^2$ gravity law but takes the mass of the convection zone into account. The difference between this approach and and the one polytropic model from \cite{jones2011}, that is implemented by default in the Rayleigh code, is negligible for shallow convection zones but becomes noticeable for deep convection zones. Our approach provides a valid approximation as long as the luminosity is constant throughout the convection zone and thus allows the treatment of K giants as well as low mass main sequence stars. 

\begin{figure*}
\begin{center}
\hspace{-1.0cm}
\includegraphics[width=8.0cm]{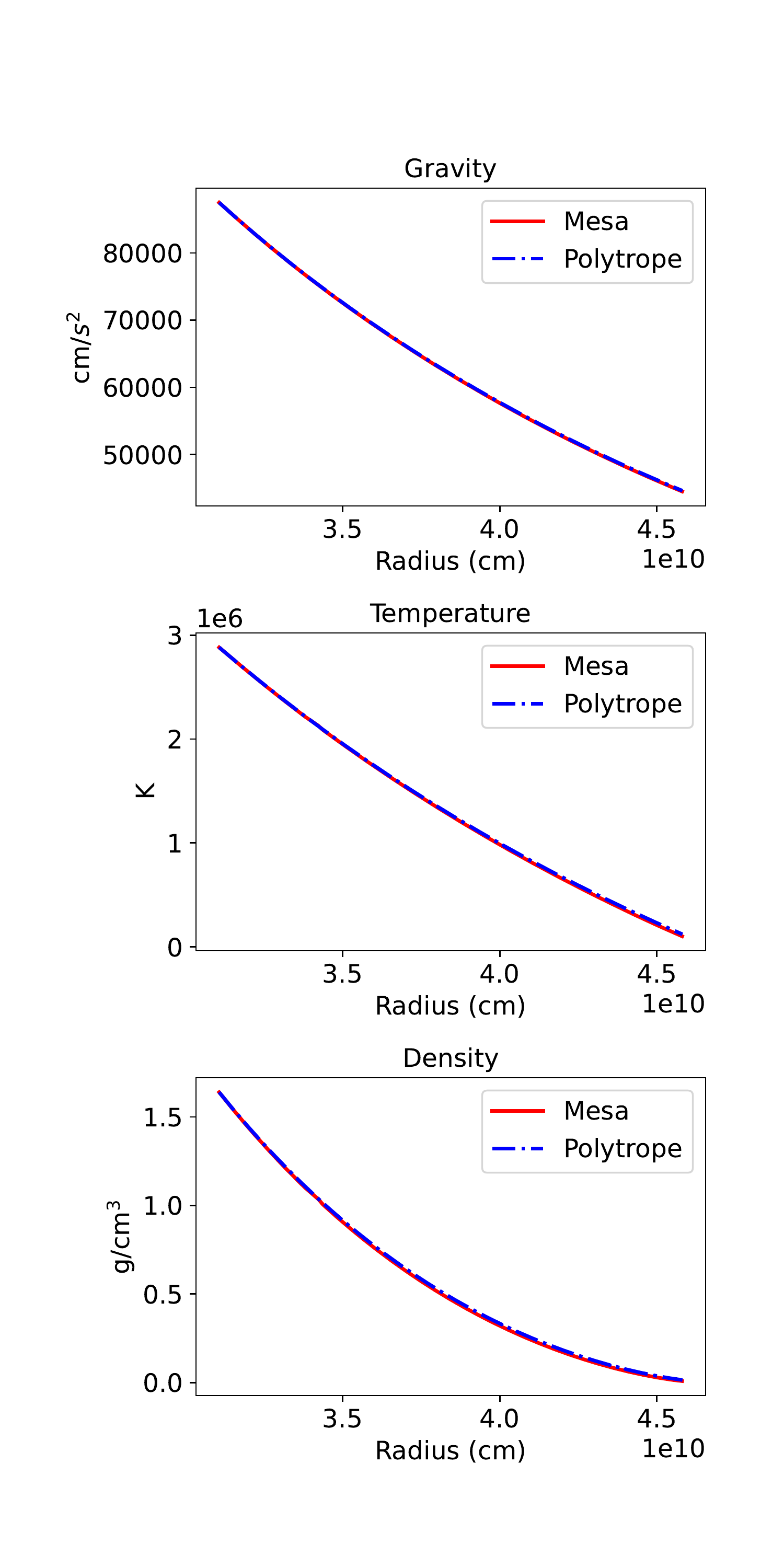}
\includegraphics[width=8.0cm]{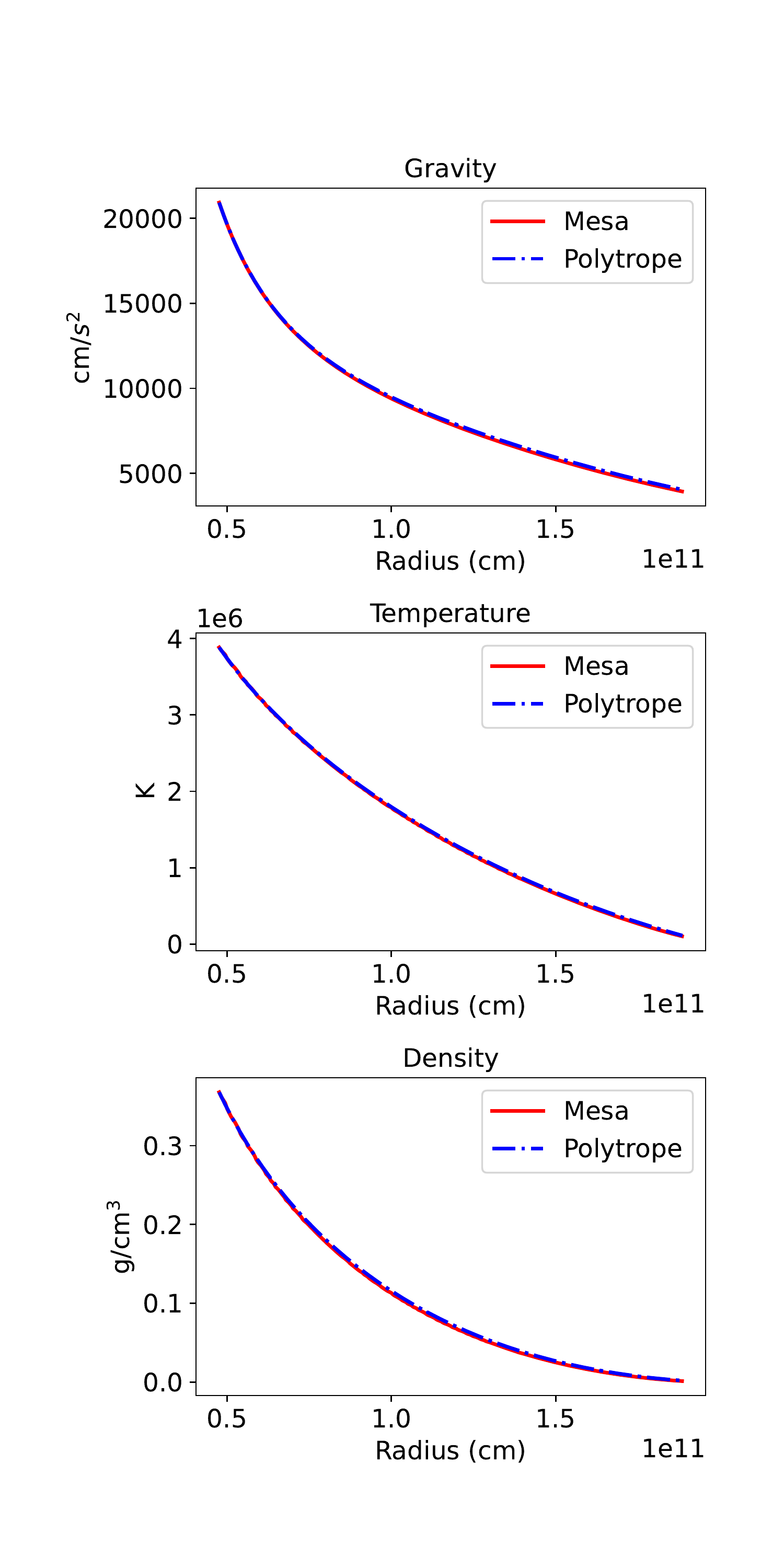}
\end{center}
\caption{Stratification in the outer convection zone of the $0.7 M_\odot$ K dwarf (left) and a $1.0 M_\odot$ K giant. The solid red lines shows the output from the {\sc Mesa} code, the dash-dotted blue lines our polytropic approximation.
\label{strat}}
\end{figure*}

Figure \ref{strat} shows the gravity force, temperature, and mass density in the convection zone of the star from the {\sc Mesa} code versus the same quantities from the polytropic model for our $0.7 M_\odot$ K dwarf and a $1.0 M_\odot$ K giant. The difference between the {\sc Mesa} and polytropic models is less than the line width for all three quantities in both cases. This shows that the polytropic model is indeed an excellent representation of the {\sc Mesa} models for giants as well as main sequence stars.

The steep stratfication at the top of the convection zone prevents the inclusion of that layer in the simulations as the Mach number is no longer small, as required for the anelastic approximation to be valid. Moreover, the small scale height in that layer would require a prohibitively high numerical resolution.
The upper boundary of our model is therefore located at a fractional radius of 0.94. This limits the range of mass densities to three scale heights. Convection is driven by a volume heating term in the lower part of the shell. We start with a small perturbation which triggers the convective instability. The system eventually settles in a state where the heat flux through the shell is constant and predominantly convective. The simulations include a magnetic field, which also starts as a small perturbation.  
Like the velocity field, the magnetic field grows exponentially until it saturates. 

\begin{figure}
\begin{center}
\includegraphics[width=9.5cm]{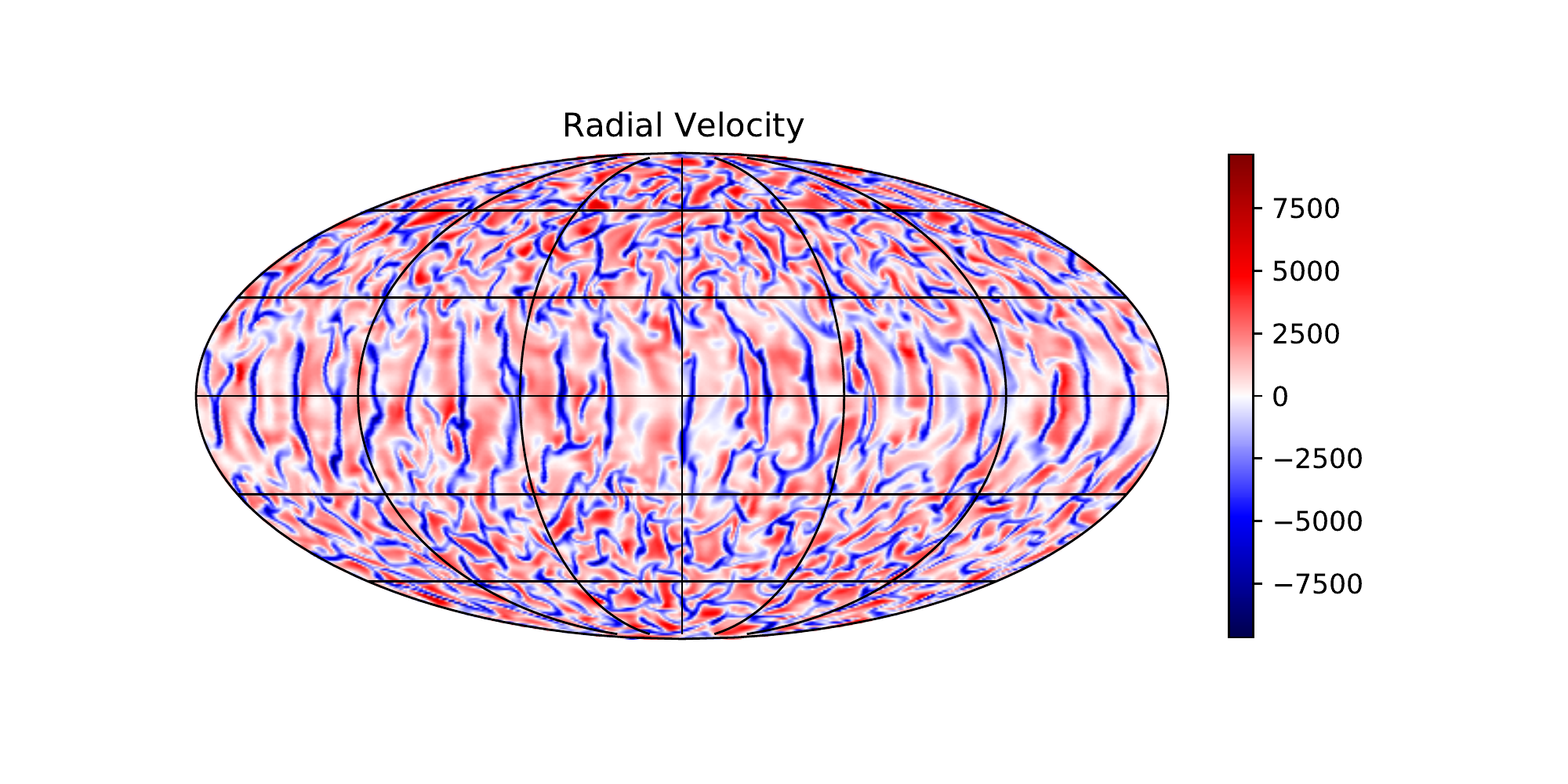}\\
\includegraphics[width=9.5cm]{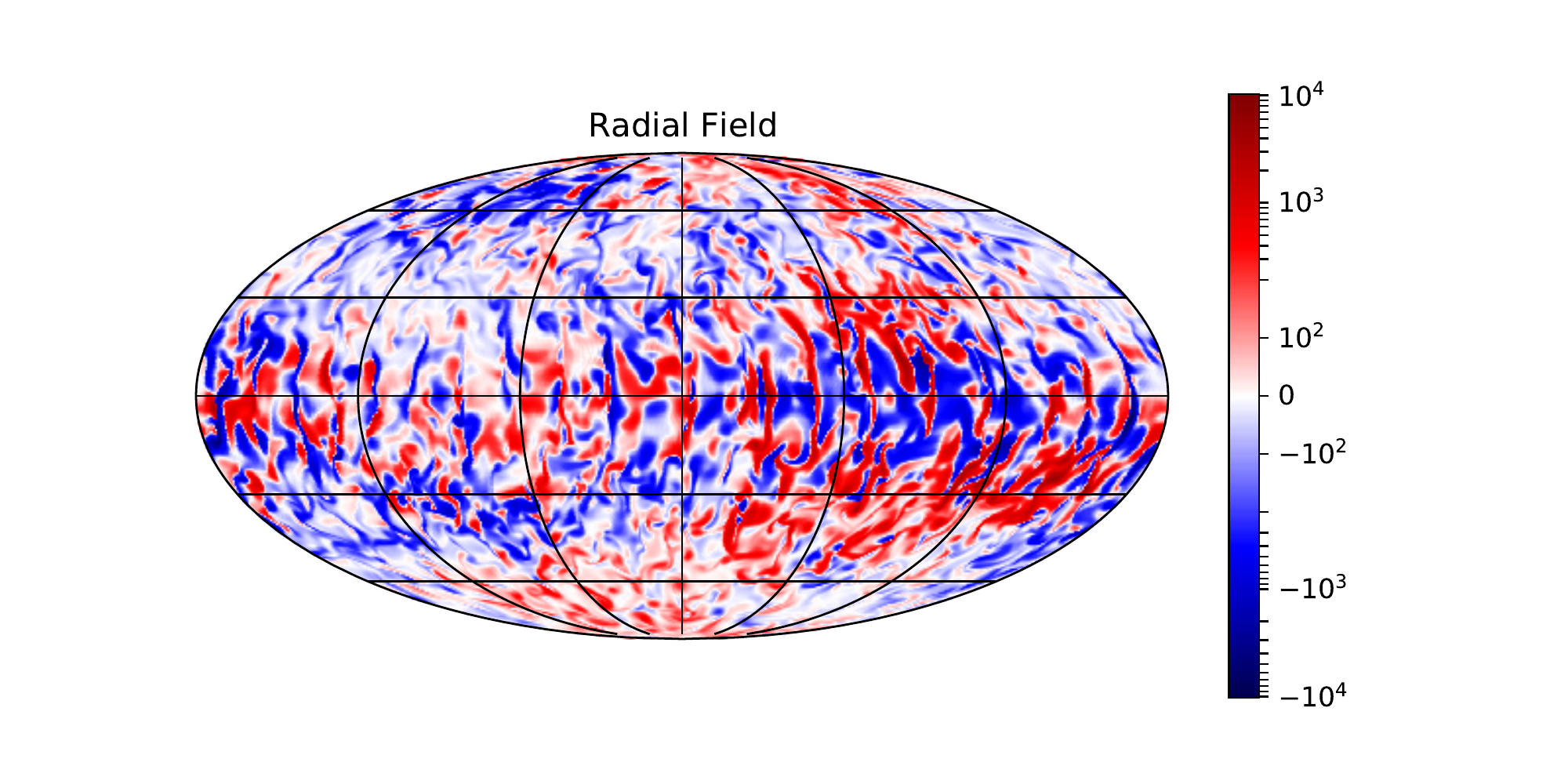}
\end{center}
\caption{Mollweide projection of surface radial velocity (top) and magnetic field components (bottom) from a model with $\Omega=2.7\times10^{-6} {\rm s}^{-1}$. 
\label{moll}}
\end{figure}

\begin{figure}
\begin{center}
\includegraphics[width=9.5cm]{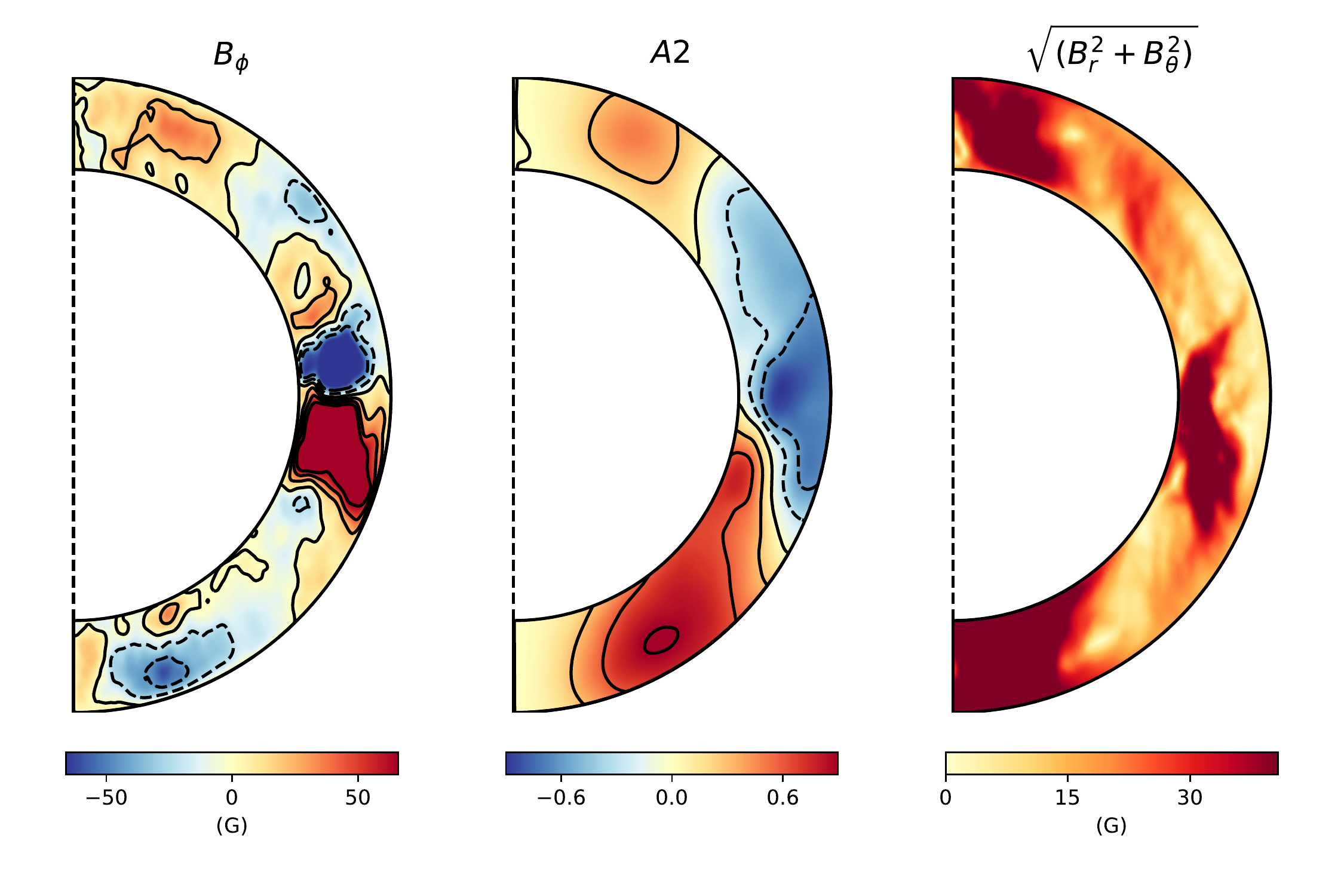}
\end{center}
\caption{Left: Azimuthally averaged values of the azimuthal magnetic field component. Center: Stream function for the azimuthally averaged poloidal field components. Right: amplitude of the poloidal field.
\label{azmag} 
}
\end{figure}

Figure \ref{moll} shows a snapshot of the radial velocity and magnetic field components at the upper boundary for a K dwarf rotating with the solar rotation period. This corresponds to a value of 10.8 for the Coriolis number and therefore represents a case of fairly, but not extremely rapid rotation. Correspondingly, the convection pattern at low latitudes shows columns that are aligned with the rotation axis rather than a B\'enard cell pattern found for slow rotation. At high latitudes, the cell structure is more plume-like. 
With field strength in the kG range, the surface  magnetic field is quite strong and clearly shows some large-scale pattern but no simple dipole geometry. Likewise, an axisymmetric component is present but weaker than the non-axisymmetric part. It is shown in Figure \ref{azmag}, which has been generated by averaging over longitude and a time span that is short relative to the total simulation time. With field strengths below 100 G, it is weaker than the fluctuating part.

\begin{figure}
\begin{center}
 \includegraphics[width=9.5cm]{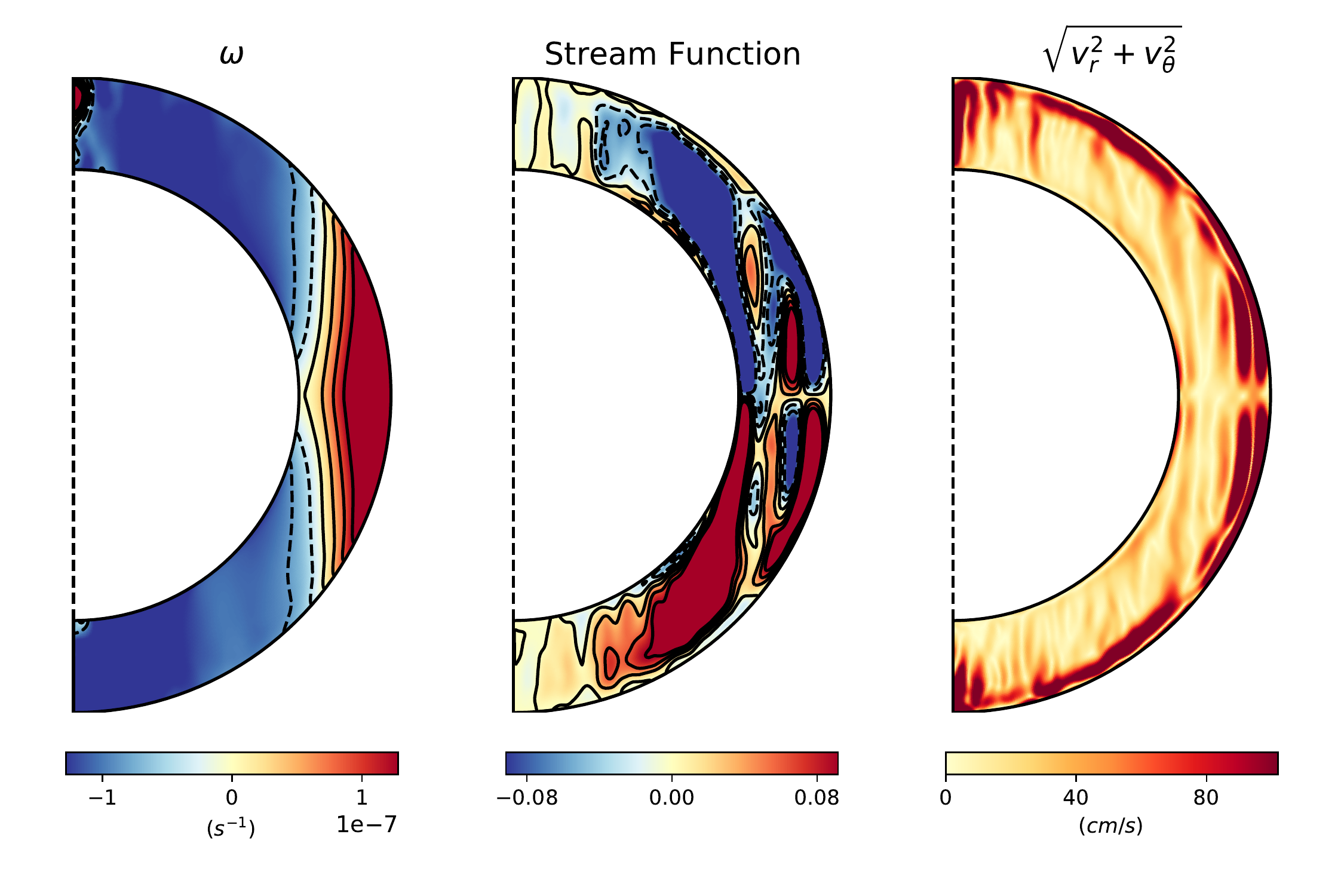}
\end{center}
\caption{Left: Azimuthally averaged values of the angular velocity. Middle: Stream function of the azimuthally averaged meridional flow components. Right: amplitude of the meridional flow.
\label{azhydro}
}
\end{figure}

Figure \ref{azhydro} shows the corresponding rotation and meridional flow patterns, which result from the same averaging procedure. The differential rotation is solar-type, i.e. the equator rotates more rapidly than the polar caps. This is in line with the findings of \cite{brun2022}, who studied a similar model, though in a slightly different parameter regime, and the general finding of anti-solar rotation for slow but solar-type differential rotation for rapid rotation \citep{gastine2014, kapyla2014, featherstone2015, mabuchi2015, viviani2018}.
A multi-cell meridional flow pattern is present at low latitudes, with amplitudes up to 10 m/s near the surface. As a large-scale flow with a single cell per hemisphere would require a significant deviation of the rotation pattern from cylindrical geometry, which we find for slow rotation only. 
%
%

\begin{figure}
\begin{center}
  \includegraphics[width=9cm]{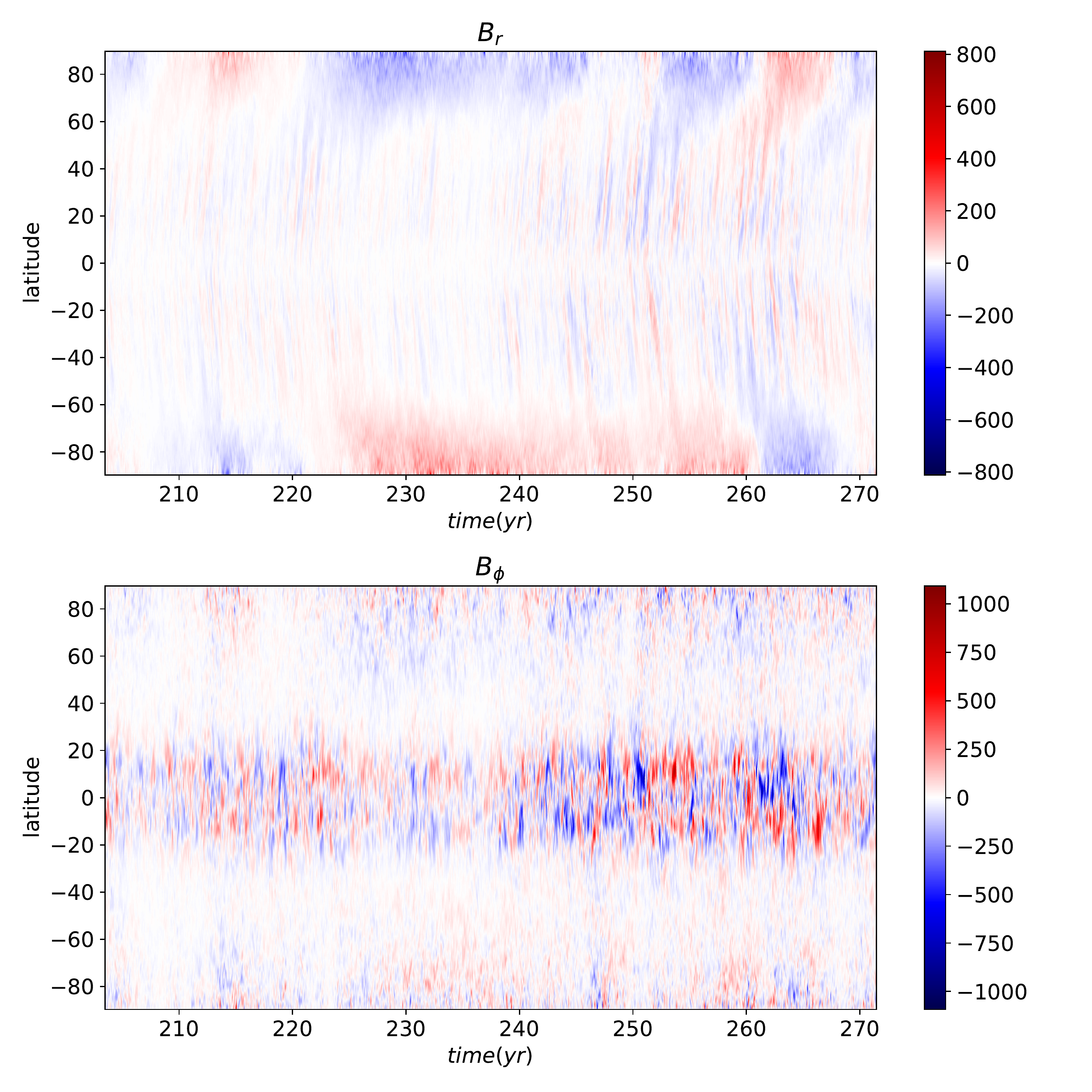}
\end{center}
\caption{Time-latitude diagram of the magnetic field. Top: the radial magnetic field near the stellar surface. Bottom: the azimuthal field component near the bottom of the convection zone.
\label{butter}}
\end{figure}

Figure \ref{butter} shows the time-latitude diagrams of the azimuthally averaged magnetic field. The top panel shows the radial field component at the top, the lower panel the azimuthal field at the bottom of the convection zone. At high latitudes there is always one dominant polarity, with reversals on a time scale of decades and no regular cycle. 
At low latitudes up to 20 degrees there is belt of increased field strength. The field polarity at any particular latitude changes much more often than at high latitudes. 
Both the high and low latitude regions show a distinct asymmetry with respect to the equator, though no perfect anisymmetry.
\section{Conclusions}
Our polytropic model is a very good approximation of the stratification from the {\sc Mesa} code for cases in which the mass of the convection zone can not be neglected. Our simulations of a main sequence K dwarf show solar-type differential rotation, a weak mean field, but a strong small-scale surface magnetic field. For the solar rotation period used here, no activity cycle is found, though the magnetic field clearly varies with time. As the star is considerably less luminous than the Sun, the convective gas motions are slower nad the turnover times longer. This results in a Coriolis number that is larger than the solar value for the same rotation rate. We therefore find solar type rotation, while similar models for the Sun usually require higher rotation rates to reproduce the observed pattern. The latter is a known problem, which may be solved by extremely high numerical solution \citep{hotta2021}. 
\bibliographystyle{cs21proc}
\bibliography{kueker.bib}

\end{document}